\documentclass[12pt,twocolumn,aps,prb,preprint]{revtex4}   

\usepackage{amsmath}    
\usepackage{amsfonts}  
\usepackage{graphicx}   

\begin{document}

\title{On the theory of flat spacetime}
\author{M. Sultan Parvez}
 \affiliation{Louisiana State University at Alexandria, Department of 
Math and Physical Sciences, Alexandria, LA 71302}
 \email{sparvez@lsua.edu}   

\begin{abstract}
Special relativity turns out to be more than coordinate transformations 
in which the constancy of the speed of light plays the central role 
between two inertial reference frames.  
Special relativity, in essence, is a theory of four-dimensional flat 
spacetime. Euclidian space spans three of the spacetime's dimensions 
and time spans the fourth.  Properties of light may not be needed to describe spacetime, which exists independently of light.  The article shows that 
a theory of spacetime can be constructed from a geometric viewpoint 
in which the speed of light does not play any role.  Moreover postulating 
four-dimensional geometry significantly simplifies the concept of special relativity.
\end{abstract}

\maketitle

\section{Introduction}
Einstein published his famous paper\cite{E1} on the 
theory of special relativity over one hundred years ago.  Although there 
has been much work on this field of study, discussion and debate 
remains alive.

Special relativity (SR) was born out of conflicts in the late 19th century between Maxwell's electrodynamics and Newtonian mechanics.\cite{Whittaker}  
Solutions to Maxwell's 
equations in free space are electromagnetic waves with a fixed speed,
$
    c=1/\sqrt{\varepsilon_0\mu_0}
$,
which obviously contradicted with the requirement of a medium for a wave and Galilean transformations of Newtonian mechanics.  

At first, physicists thought that the problem was with Maxwell's equations. 
Attempts to modify Maxwell's equations to make them invariant under the 
Galilean transformations, however, ended in failure.  Such modifications led to 
the prediction of new electrical phenomena, which could not be found experimentally.

Then in 1903 Lorentz  made a remarkable discovery. He found that when 
he applied the Lorentz transformation of coordinates and time 
(Equation~(\ref{eq:LT})), to Maxwell's equations, the equations remained invariant. 

Maxwell's equations were only thought to express electromagnetism in the rest 
frame of ether, the postulated medium for light.  When the Michelson-Morley experiment produced a null result for the change of the velocity of light due 
to the Earth's hypothesized motion through the ether, an explanation 
was provided by Lorentz and others that ether remains undetected because  
length in the direction of motion is contracted and time is dilated. 
This apparently solved the conflicts between Maxwell's electrodynamics and 
Newtonian mechanics, but the physical meaning of the origin of the Lorentz 
transformations remained unexplained.  

In 1905 Einstein sought an alternate explanation.  Einstein's paper 
{\it On The Electrodynamics of Moving Bodies\/}\cite{Einstein1} demonstrated   
that two simple postulates, viz., (i) the equivalence of inertial frames, 
and  (ii) the invariance of the speed of light, can resolve all the asymmetries that arise
when Maxwell's electrodynamics are applied to moving objects.  The paper  
showed that an introduction of ether is 
unnecessary for the transformation of the set of coordinates of an event 
into another inertial frame's set of coordinates in a consistent theory 
of the electrodynamics of moving bodies.  The absence of ether also conformed 
with the result that ether cannot be detected experimentally.  Although Lorentz transformations, time dilation, and length contraction are derived 
in detail, the paper did not mention 
anything about the union of space and time.  Einstein used the constant speed light signal 
as the sole means of communication in all experiments and the sole means of 
interpretation of all gedanken experiments.

Minkowski\cite{Einstein1}, in 1908, greatly enhanced Einstein's work.
The most significant feature of his paper is the existence of the invariance of four-dimensional interval and the invariance of other physical quantities in 4-dimensional form.  
This new formulation gave rise to the 
concept of spacetime as we know today.  In this new formulation the Lorentz 
Transformation appears to be merely a rotation in the complex co-ordinate 
system ($t, x, y, z$).

At first Einstein was skeptical of Minkowski's work.  Einstein believed 
that Minkowski was merely rewriting the laws of special relativity in 
a new mathematical language and that the 
abstract mathematics obscured the underlying physics. 
However, Einstein changed his mind and used 4-dimensional spacetime to 
develop the general theory of relativity.

Although 4D spacetime is the underlying foundation of the general 
theory of relativity, 
SR is customarily introduced with Einstein's postulates as the 
consequences of the measurements of the constant speed of light   
without referring to spacetime geometry.  For these reasons 
novices of SR often 
incorrectly conclude that the finite speed of light creates an 
illusion in the measurement of distance and time, and that if 
signals other than light can be used as the means of communication, 
the Lorentz contraction and time dilation would disappear.  

In general relativity the bending of light rays was explained by the curvature of space-time.  
On the other hand, the second postulate in SR lacks a logical foundation, 
unlike the self-explanatory first postulate.  
Although constancy of speed of light is supported by experiments 
(e.g., the Michelson-Morley experiment), it was not supported  
by sound logical arguments. 
The uncomfortableness with the second postulate is almost as old as the 
postulate itself.\cite{tolman, pauli} 

A student of SR rightfully suspects the result of Michaelson-Morley experiment 
and its interpretation.  Certainly an experiment can alter the result of a 
previous experiment, and interpretations of experiments can also differ.

We attempt here to reconstruct 
a version of SR using a geometrical point of view of spacetime without 
using light.  If spacetime exists independent of light, 
a description of spacetime should not depend exclusively on light.

\section{The New Postulate}
Obviously the first thing to do to remove the role played by the light 
in SR is to replace the 
second postulate.  This can be done by an alternate postulate as follows: Space and time form 
a 4-dimensional continuum.  

Furthermore,  
both the postulates of SR can be replaced by one: 
{\it Spacetime is a four-dimensional 
continuum with time being the fourth dimension extension of 
3D Euclidian space, and all nonaccelerating reference frames are 
equivalent.}  
The association of light is eliminated from the postulates of 
SR and is replaced with spacetime geometry.

Since the proposed postulate is an extension of 3-D space, 
it is logically comprehensible and is less intrusive as 
postulating the original second postulate.  The space 
coordinates and time are also found to have 
the same status in the wave equations in mechanics and 
electrodynamics.

\section{DEPICTION OF SPACETIME AND SCALING THE TIME AXIS}
Historically time and space are measured in two different units.\cite{Grav1}  Most importantly we have to use two different types of instruments to measure 
them. A stationary ruler cannot be used to measure time, nor can a clock 
be used to measure distance.   Our postulate of 4D spacetime demands that 
the time measurement and the distance measurement must be related.  
Drawing a coordinate system of the spacetime 
could resolve this and many other issues.
Since we proposed that time is another dimension in spacetime, 
we have a situation in which time can be represented by an axis 
perpendicular to all three space axes.

A 2D-slice of the spacetime can be depicted, as is done customarily, by plotting a reference frame with time, $t$, along the vertical and $x$ axis only of the space along the horizontal.  Similarly a diagram of 3D-slice of the spacetime can also be constructed by adding the $y$ axis along the horizontal.  A diagram of the whole spacetime, however, is impossible.

Again, as is customary, a point in the spacetime diagram represents an 
{\it event,} and a line represents a {\it world line\/} of a particle 
moving with a speed $v$, where,
\begin{equation}
v = {dx \over dt} = {1 \over dt/dx} = {1 \over \mbox{slope of the line}} \nonumber
\end{equation} 

In spacetime diagram, time is another dimension.  How are time and space scales related in a spacetime diagram?  
The time unit and distance unit must be proportional 
to each other,\cite{STPhys1} i.e., 
\begin{equation}
|\Delta\hat{t}| = {1 \over c}\; |\Delta \hat{r}| \label{trhat}
\end{equation}
where $\Delta \hat{t}$ is the unit time separation in the 4D spacetime and $\Delta \hat{r}$ is a unit distance in the 3D Euclidian space. $c$ is the constant of proportionality between space and 
time units, and is also known as the spacetime conversion factor.  
Equation~(\ref{trhat}) should be valid in all nonaccelerating reference 
frames; otherwise, the reference frames would not be equivalent.  If 
the value of $c$ depends on the reference frame, then the reference 
frames are not equivalent.  Equation~(\ref{trhat}) is the universal 
relation between time and space.  $c$ is a frame independent universal constant.


Equation~(\ref{trhat}) can be 
used to define a new unit of time in terms of a conventional unit of distance (or it can also be  used to define a new unit of distance in terms of a conventional unit of time) : 
\begin{equation}
|\Delta\hat{t}| = {1 \over c}\times 1.0 \text{ meter}\label{tdef}
\end{equation}
It is customary in this type of situation to set the conversion factor or 
the proportionality constant $c$ to 1 to make 
Equation~(\ref{tdef}) the defining equation for a new unit of 
time.  Since we have chosen $c$ to be unitless, this new unit of time is {\it meter.}

Using the relation, $ (\Delta\hat{r})^2 =(\Delta x_{\hat{r}})^2 + (\Delta y_{\hat{r}})^2 
+ (\Delta z_{\hat{r}})^2$ of 
Euclidian geometry, we get from Equation~(\ref{trhat}),
\begin{equation} 
(\Delta\hat{t})^2 = (\Delta x_{\hat{r}})^2 + (\Delta y_{\hat{r}})^2
+ (\Delta z_{\hat{r}})^2  \label{txyzhat}
\end{equation}
assuming space is isotropic.   $\Delta x_{\hat{r}}$, $\Delta y_{\hat{r}}$, and $\Delta x_{\hat{r}}$ 
are the magnitudes of three components of a unit vector 
$\Delta\bf{\hat{r}}$.  There is a set 
of values for $\Delta x_{\hat{r}}$, $\Delta y_{\hat{r}}$, and $\Delta x_{\hat{r}}$ to satisfy 
Equation~(\ref{txyzhat}). Similarly in another inertial frame, 
S$^\prime$, moving with some speed with respect to frame S,
\begin{equation} 
(\Delta\hat{t^\prime})^2 = 
(\Delta x^\prime_{\hat{r}})^2 
+ (\Delta y^\prime_{\hat{r}})^2 
+ (\Delta z^\prime_{\hat{r}})^2  \label{txyzprimehat}
\end{equation}

Since it is impossible to draw and work with a 4D diagram, it is instructive to work with the special case of the 2D slice---the \mbox{$t$-$x$} plane---of the 4D-spacetime and then generalize the work to the 4D case.  In the  $t$-$x$ plane 
Equation~(\ref{txyzhat}) reduces to 
\begin{equation}
|\Delta\hat{t}| = |\Delta\hat{x}| \label{txhat}
\end{equation}

Similarly in the frame S$^\prime$, we get from Equation~(\ref{txyzprimehat}),
\begin{equation}
|\Delta\hat{t^\prime}| = |\Delta\hat{x^\prime}| \label{txprimehat}
\end{equation}

In actual spacetime there is no coordinate system, let alone a prefered 
coordinate system.  There are only events.  To calibrate their time scale, 
observers must rely on special events whose time and space separations are 
identified as equal.  That means there are two things the observer has 
to consider: 
1) The same set of events must be used by all observers for calibration of 
their time axis or to determine the spacetime conversion factor; and 
2) These 
events must be a special set of events, such that the ratio of the space separation to the time separation of any of these events is same in all 
frames. 

The scaling equations, Equations~(\ref{txhat}) and (\ref{txprimehat}), 
tell 
more than just how to plot the time scale compared to the space scale.  An 
equivalent expression to Equation~(\ref{txhat}) can be written as
\begin{equation} 
\Delta t_0 = \Delta x_0 \label{t0x0}
\end{equation}
where $\Delta t_0$ can be written as $|\Delta\hat{t}|s$, $s$ is a real number.
Similarly $\Delta x_0$ can be written as $|\Delta\hat{x}|s$.  Only a set of 
events' time and space coordinates in a reference frame satisfy 
Equation~(\ref{t0x0}).  In the $t$-$x$ plane the locus of those special events 
is the straight line making 45$^\circ$ with $x$-axis.  The special events 
that satisfy Equation~(\ref{t0x0}) can be used to calibrate a clock or to 
determine the spacetime conversion factor $c$.  Observers must agree 
on the special events and all observers must use the same events to calibrate 
their time axis or determine the spacetime conversion factor.  This is the 
only way the time axis can be calibrated.  Otherwise, if two observers have 
their own set of special events that differ and calibrate their  
time axes differently or determine different values of the spacetime 
conversion factor, then the transformation of coordinates of an event 
from one frame to another would be meaningless.  If an observer 
identifies an event as a special event, it must also be a special 
event in all other frames.  

The scaling equation also describes two other types of events:  (1) 
the events 
for which their time coordinates $\Delta t$ are greater than their 
space coordinates $\Delta x$, known as timelike events, and (2) the 
events 
for which their space coordinates $\Delta x$ are greater than their 
time coordinates $\Delta t$, known as spacelike events.  Although the 
special events are otherwise known as lightlike events, we will 
call them special events here and put a subscript ``$_0$'' to 
distinguish them from other events.  

The implication of Equation~(\ref{t0x0}) is that two events that 
have equal time and space separations will also have equal 
time and space separations in all other inertial reference frames.
That is,
\begin{equation}
\Delta t_0 = \Delta x_0 = \Delta t^\prime_0 = \Delta x^\prime_0 
\label{t0x0t'0x'0}
\end{equation}

Measuring from the origin, (Measuring coordinates of a given event means 
measuring with respect to an event at the origin.  Referencing  
coordinates of a given event always implies dealing with two events---the 
given event and the event origin.) the set of special events  
$\{\Delta t_0, \Delta x_0\}$ that satisfies 
Equation~(\ref{t0x0}) have equal time and space separation and forms a straight line in the $t$-$x$ plane with slope 1 (see Figure 1).  
We will call this line the {\it special line.}
\begin{figure}[h]
 \setlength{\unitlength}{0.5in}
\begin{picture}(5,5)(-.5,0)

\put(0,0){\line(1,0){5}}
\put(0,0){\line(0,1){5}}

\put(0,0){\line(1,1){4.5}}
\put(3,2.5){\large$(t_0\,,x_0)$}
\put(2.6,2.6){\circle*{.1}}

\put(3,-.3){\large$x$}
\put(-.3,3){\large$t$}

\end{picture}
 \caption{\label{Fig1}The $t-x$ plane of the spacetime diagram showing 
the special line and a special event.}
\end{figure}
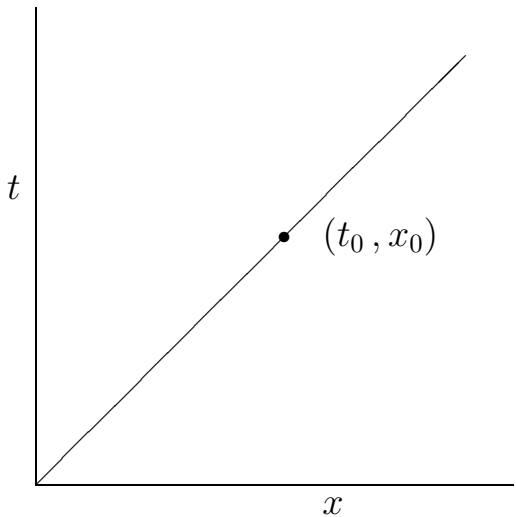

Also note that the set of special events  $\{\Delta t_0, \Delta x_0\}$ 
is the locus of the coordinates of a moving object with a velocity of 
one (or $c$ in conventional unit).  
Such objects will also move with the same speed in all other reference 
frames.  Therefore a speed of 1 is an invariant speed.

Suppose an observer S uses the coordinates $t$ and $x$ as above and that another
observer S$^\prime$, with coordinates $t^\prime$ and $x^\prime$ is moving 
along the 
positive $x$ direction of S with a velocity $v$.  Suppose again that 
the two events {\it 1\/} and {\it 2\/} are recorded 
as $(t^\prime_0,\, 0)$ and $(0,\, x^\prime_0)$, respectively  
in frame S$^\prime$.  The subscript $0$ again means that the time 
separation and the space separation between the two 
events are same, i.e.,
\begin{equation}
|0-t^\prime_0| = |x^\prime_0 -0| \label{0t0xprime}
\end{equation}

How do these two events look from S?

The $t^\prime$ axis is the world line of a particle at rest at 
$x^\prime$ = 0 with respect to S$^\prime$. The world line of the same particle in S 
will be a straight line with a slope of 1/$v$.  The $t^\prime$ axis looks like that shown in Figure~2.  Event {\it 1\/} is somewhere on 
the $t^\prime$ axis as in the Figure~2 with coordinates 
$(t_{10}, x_{10})$.
\begin{figure}[h]
 \setlength{\unitlength}{0.5in}
\begin{picture}(5,5)(-.5,0)

\put(0,0){\line(1,0){5}}
\put(0,0){\line(0,1){5}}
\put(3,-.3){\large$x$}
\put(-.3,3){\large$t$}
\put(4.6,1.1){\large$x^\prime$}
\put(1.1,4.6){\large$t^\prime$}

\put(0,0){\line(1,4){1.1}}
\put(0,0){\line(4,1){4.5}}
\put(.9,2.5){\large$(t_{10}\,,x_{10})$}
\put(2.8,.4){\large$(t_{20}\,,x_{20})$}
\put(.63,2.5){\circle*{.1}}
\put(2.5,.63){\circle*{.1}}
\put(2.5,0){\dashbox{.1}(0,.63)}
\put(0,.63){\dashbox{.1}(2.5,0)}
\put(0,2.5){\dashbox{.1}(.63,0)}
\put(.63,0){\dashbox{.1}(0,2.5)}

\put(1.5,.8){$\Delta x_0$}
\put(1.4,.85){\vector(-1,0){.75}}
\put(2.1,.85){\vector(1,0){.35}}

\put(.8,1.5){$\Delta t_0$}
\put(.9,1.7){\vector(0,1){.75}}
\put(.9,1.4){\vector(0,-1){.75}}

\end{picture}
 \caption{\label{Fig2}The $t^\prime$ and the $x^\prime$-axes 
of the frame S$^\prime$ as observed from the frame S when 
both origins coincide.  The S$^\prime$ is moving with a speed $v$ 
in the positive $x$ direction of S (the standard configuration).}
\end{figure}
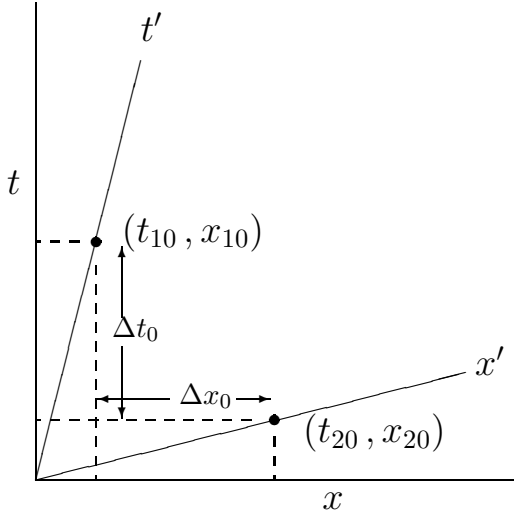

Since events {\it 1\/} and {\it 2\/} satisfy Equation~(\ref{0t0xprime}) in 
frame S$^\prime$, in frame S they will also satisfy,
\begin{equation}
|t_{20} - t_{10}| = |x_{20}-x_{10}| \label{t2-t1,x2-x1}
\end{equation}

From the point of view of S, $\Delta t_0$ represents the measurement 
of the time interval between events $(t_{20},\,x_{20})$ and 
$(t_{10},\,x_{10})$, and 
$\Delta x_0$ represents the space interval between the same two events.  
Event {\it 2\/} $(t_{20},\,x_{20})$ which satisfies Equation~(\ref{t2-t1,x2-x1}) is 
shown in Figure~2. To keep 
the value of spacetime conversion factor the same and the time and space 
interval equal in the frame S,  the $x^\prime$-axis must be a 
line with a slope of $v$ in the frame S.  It is clear that the simultaneous events ($x^\prime$-axis) in frame S$^\prime$ are not simultaneous in  frame S.


\section{THE SPACETIME METRIC}
Our postulate that space and time form a 4D continuum demands that 
space-time separations must have an invariant relation under changes of 
coordinates.  Otherwise space and time would be two separate entities.  
As a matter of fact, the existence of a spacetime metric equation 
is the necessary and sufficient condition for the spacetime to be a 
4D continuum.

We assign three dimensions $(x,y,z)$ to space because the distance 
$ d^2 = \Delta x^2 + \Delta y^2 + \Delta z^2 $ between two points in 
space is invariant under rotation or translation of coordinates.  
We know that the 4D Euclidean metric equation, 
$ \Delta d_4^2 = \Delta t^2 + \Delta x^2 + \Delta y^2 + \Delta z^2 $, 
between two events in spacetime does not remain invariant under 
changes of coordinates between two inertial reference frames.  

Consider two inertial reference frames, S with coordinates 
 ($t$, $x$, $y$, $z$) and S$^\prime$ 
with coordinates ($t^\prime$, $x^\prime$, $y^\prime$, $z^\prime$), with S$^\prime$ moving at a 
constant speed $v$ relative to S  along the direction of the 
positive $x$ axis.  Let $x$ and $x^\prime$ axes coincide along the 
direction of relative motion.  Let the coincidence of the origins of 
S and S$^\prime$ be event {\it 1}.  Event {\it 1\/} 
can be recorded in both frames as  ($t_1$ = $x_1$ = $y_1$ = $z_1$ = 0) and 
($t_1^\prime$ = $x_1^\prime$ = $y_1^\prime$ = $z_1^\prime$ = 0)  
according to the standard configuration. Suppose another event {\it 2\/} 
is recorded in frame S$^\prime$ as ($t_2^\prime$, 0, 0, 0) as shown 
in the Figure~3.  How would the time separation $\Delta t^\prime$ 
= $t_2^\prime - t_1^\prime$ = $t_2^\prime - 0$ be recorded in the frame S?  
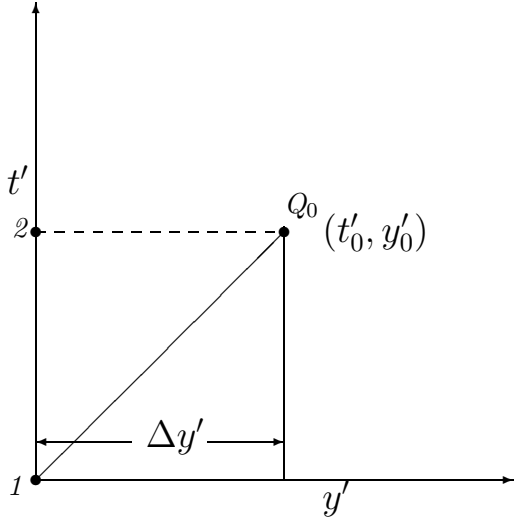
\begin{figure}[h]
 \setlength{\unitlength}{0.5in}
\begin{picture}(5,5)(-.5,0)

\put(0,0){\vector(1,0){5}} 
\put(0,0){\vector(0,1){5}}

\put(0,0){\line(1,1){2.6}}
\put(0,2.6){\dashbox{.1}(2.6,0)}
\put(2.6,0){\line(0,1){2.6}}

\put(2.6,2.8){\it Q$_0$}
\put(3,2.5){\large$(t^\prime_0, y_0^\prime)$}
\put(0,2.6){\circle*{.1}}
\put(-.25,2.55){\it 2}
\put(2.6,2.6){\circle*{.1}}
\put(0,0){\circle*{.1}}
\put(-.3,-.15){\it 1}

\put(3,-.3){\large$y^\prime$}
\put(-.3,3){\large$t^\prime$}

\put(1,.4){\vector(-1,0){1}}
\put(1.15,.35){\large$\Delta y^\prime$}
\put(1.8,.4){\vector(1,0){0.8}}

\end{picture}
 \caption{\label{Fig3}Events {\it 1, 2\/} and {\it Q}$_0$ in frame 
S$^\prime$. {\it Q}$_0$ is in the $t^\prime$-$y^\prime$ 
plane.}
\end{figure}

We shall use the properties of the special coordinates of spacetime 
and of the Pythagorean relation of space coordinates to examine how the separation between two events transforms in another frame of reference.

Consider a special event {\it Q$_0$}($t_0^\prime$, 0, $y_0^\prime$, 0) 
simultaneous to event {\it 2\/} in frame S$^\prime$ as shown in the 
Figure~3. Because {\it Q$_0$} is a spacial event, 
\begin{equation}
\Delta t^\prime = t_0^\prime = y_0^\prime = \Delta y^\prime 
\label{tprimeyprime}
\end{equation}
where $\Delta y^\prime$ = $y_0^\prime - y_1^\prime$, and, of course 
$y_1^\prime$ = 0.

Now let us look at the events {\it 1}, {\it 2\/} and {\it Q$_0$} in the spacetime 
diagram of S.  Event {\it 1\/} is on the origin of 
frame S.
Event {\it 2} is on the $t^\prime$ axis 
and will have coordinates ($t_2$, $x_2$ = $vt_2$, 0, 0) as shown in the 
Figure~4.  
\begin{figure}[h]
 \resizebox{!}{2.3in}{\includegraphics{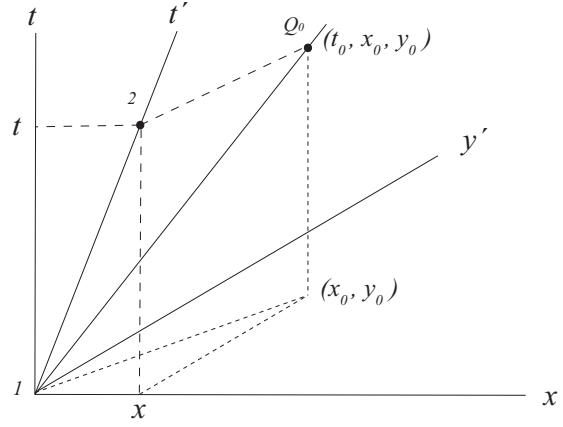}}
 \caption{\label{Fig4}Events {\it 1, 2, Q}$_0$ and frame S$^\prime$ in frame S. }
\end{figure}
Event {\it Q$_0$} has the coordinates ($t_0$ = $t_2$, 
$x_0$ = $vt_2$, $y_0$ = $y_0^\prime$, 0) in frame S as shown 
in the Figure~4.  Because {\it Q}$_0$ is a special event, its time separation 
and space separation are equal,
\begin{equation}
(t_0 - t_1)^2 = (x_0-x_1)^2 + (y_0-y_1)^2 \nonumber
\end{equation}
or,
\begin{equation}
\Delta t^2 = \Delta x^2 + \Delta y^2   \label{ty}
\end{equation}
where $\Delta t$ = ($t_0$ $-$ $t_1$) =  ($t_2$ $-$ $t_1$), $\Delta x$ = 
($x_0$ $-$ $x_1$), and $\Delta y$ = ($y_0$ $-$ $y_1$).  Now $\Delta y$ 
and $\Delta y^\prime$ are equal. Combining Equations~(\ref{tprimeyprime}) and 
(\ref{ty}) we get,
\begin{equation}
\Delta {t^\prime}\,^2 = \Delta t^2 - \Delta x^2
\end{equation}

Similarly in another frame S$^{\prime\prime}$,
\begin{equation}
\Delta {t^\prime}\,^2 = 
\Delta {t^{\prime\prime}}\,^2 - \Delta {x^{\prime\prime}}\,^2
\end{equation}

So, the separation of coordinates of two events measured from two 
inertial frames moving in the $x$ direction with respect to 
each other is related by, 
\begin{equation}
\Delta t^2 - \Delta x^2 = \Delta {t^\prime}\,^2 - \Delta {x^\prime}\,^2 
\label{tsqminusxsq}
\end{equation}

Generalizing Equation~(\ref{tsqminusxsq}) in 4D spacetime, the space and 
time separations between two events observed in 
two inertial frames must obey,  
\begin{eqnarray}
(\Delta t^\prime)^2 - 
(\Delta x^\prime)^2 + (\Delta y^\prime)^2 + (\Delta z^\prime)^2 \nonumber \\ 
\label{4dinterval}
= (\Delta t)^2 - (\Delta x)^2 + (\Delta y)^2 + (\Delta z)^2  
\end{eqnarray}

Equation~(\ref{4dinterval}) prompted us to define the invariant interval between 
any two events that are separated by coordinate 
increments $(\Delta t, \Delta x, \Delta y, \Delta z)$ as 
\begin{equation}
\Delta s^2 = (\Delta t)^2 - (\Delta x)^2 + (\Delta y)^2 + (\Delta z)^2  
\label{eq:interval}
\end{equation}

Equation~(\ref{eq:interval}) can be verified by calculating intervals in two 
different frames.

\section{THE LORENTZ TRANSFORMATIONS}
The Lorentz transformations can be deduced from the spacetime metric 
equation~(\ref{eq:interval}).  

The Lorentz transformations can actually be derived in a 
variety of ways.\cite{LT1, LT2, LT3, LT4} For example, the original 
derivation\cite{Lorentz1, Lorentz2}  of the transformation 
by Lorentz occured prior to the theory of special relativity and 
did not use the 
postulates of relativity at all.  Lorentz used the negative result of the 
Michelson-Morley experiment and the field equation of Maxwell to derive 
length contraction.

The derivation of the Lorentz transformation from the invariance of interval 
can also be found in many books.\cite{Schutz}   The derivation has been 
included here for a review and for the completeness of the topic.

The Lorentz transformation expresses the coordinates of S$^\prime$, which  
moves with speed $v$ on the positive $x$-axis relative to S, in terms of 
the coordinates of S. The lengths perpendicular to the $x$-axis are 
the same when measured by S or S$^\prime$.  The most general 
linear transformation, then, is
\begin{subequations}
\begin{eqnarray}
\label{eq:tbar}
t^\prime &=& \gamma t + \beta x \\
\label{eq:xbar} x^\prime &=& \alpha t + \sigma x\\ 
 y^\prime &=& y\\
 z^\prime &=& z
\end{eqnarray}
\end{subequations}
$\gamma, \beta, \alpha$, and $\sigma$ are at most a function of $v$.

From the considerations of the inertial frames alone, one can easily 
show\cite{Schutz} that 
\begin{equation}
\sigma = \gamma 
\end{equation}
and 
\begin{equation}
{\alpha\over\sigma} = {\beta\over\gamma} = -v 
\end{equation}
Therefore the transformation equations (\ref{eq:tbar}) and (\ref{eq:xbar}) become,
\begin{eqnarray}
t^\prime = \gamma(t - vx)  \label{eq:tvx}\\
x^\prime = \gamma (x - vt) \label{eq:xvt}
\end{eqnarray}

Now comes the most important part of the derivation: using the invariance of 
the interval.  Substituting Equations~(\ref{eq:tvx}) and 
(\ref{eq:xvt}) in  (\ref{eq:interval}) and after some straightforward 
calculations, one gets,
\begin{equation}
\gamma = \pm {1 \over \sqrt{1 - v^2 } }
\end{equation}
The $+$ sign is the proper choice in the above equation to avoid an inversion 
of the coordinates.

Therefore, the complete Lorentz transformations is, 
\begin{subequations}
\label{eq:LT}
\begin{eqnarray}
\label{eq:LTt}
t^\prime &=& {t-vx \over \sqrt{1-v^2} } \\
\label{eq:LTx}
x^\prime &=& {x-vt \over \sqrt{1-v^2} } \\
y^\prime &=& y \\
z^\prime &=& z  
\end{eqnarray}
\end{subequations}

\section{Determination of the spacetime conversion factor}
Determination of $c$ using any equation and method of special 
relativity---from the point of view of 4D-spacetime geometry---will 
give the value of the spacetime conversion factor and 
not the speed of light. For example if we use a $K$ meson decay data in 
laboratory frame and rest frame to determine $c$ using 
Equation~(\ref{eq:interval}),
the value of $c$ thus obtained would be the value of the spacetime conversion
factor and not of the speed of light.  

Another way to determine the value of the spacetime factor would be from 
the measurements of the speed of a moving object from two different 
reference frames and by using the velocity addition formula.  A third 
way would be using Equation~(\ref{t0x0}) and identifying the special events.
Suppose when several reference frames coincide, a firecracker explodes 
at the origin.  Suppose in one frame at a distance of 1.0 m, another
firecracker explodes at the time of 3.34 ns (1/$c$ seconds, $c$ in m/s).  
These two events 
would be special events, and the special events in one frame means 
one has 
special events in all frames.  Measuring their space and time 
coordinates and using Equation~(\ref{t0x0}), all observers can determine 
the spacetime conversion factor.  In practice,  a 
number of firecrackers must explode at different times and the set 
which gives the 
same value of the ratio of space to time coordinates would be 
identified as the special events.  

None of the above mentioned methods uses light.  Would it not be 
interesting to compare the results thus obtained with the speed of 
light?  Also, to date, the accuracy of the above methods does not match 
the accuracy of determining the speed of light.  

How do the  spacetime conversion factor and the speed of light differ?
In spacetime, light signals produce events whose time and space coordinates 
are equally separated.  All observers see them equally separated.  
Measuring the coordinates of these events and taking the ratio to 
determine the spacetime conversion factor is essentially the same 
as measuring the speed of light! Conversely, we realize that the trail 
of events produced by light are the special events from the measurement 
that the speed of light is the same in all reference frame.  
The only connection we can see between the speed of light and the spacetime 
conversion factor is that if an object's time and space coordinates satisfy 
Equation~(\ref{t0x0}), then the object is moving with the speed $c$ and 
that speed will be the same for all reference frames.
The fact that the speed of light is a constant speed for all 
observers---as measured in the Michelson-Morley experiment---confirms 
that light produces special events.  Therefore, the speed of light 
and the spacetime conversion factor should not have different values.

\section{RESULTS AND DISCUSSION}
According to Einstein's (the traditional view of the 
special relativity theory) point of view, the constancy of the speed 
of light has the status of a law of nature, and the Lorentz 
transformations are required to keep the speed of light constant.  
Space contraction, time dilation, and time desynchronization follow 
as a logical necessity from the empirical fact of the constancy of 
the speed of light.  There is no way to reason from the knowledge 
of space and time itself that space contraction and time dilation and 
desynchronization can take place.  

Spacetime geometry, on the other hand, can explain all the counter 
intuitive notions of special relativity including the constancy 
of the speed of light.  First of all, there is no prefered 
reference frame because there are no reference frames in spacetime, 
but rather only events.  We artificially construct reference 
frames for the convenience of making measurements and keeping records, 
and equivalent reference frames are the best we can construct!  
There are three 
types of events.  Events that are time-like are time-like to 
all equivalent observers.  Events that are space-like are 
space-like to all equivalent observers.  Events that are 
special are special to all equivalent observers.

It is the defective construction of the reference frames that 
creates the ``illusion'' of the length contraction, time 
dilation, time desynchronization, and a finite 
constant speed of light in our measurements.  
If a reference frame measures 
the proper time between two events, that reference frame 
cannot measure the proper distance between those events.  
Similarly an observer who can measure the proper distance 
between two events cannot measure the proper time between 
those events.  Consider the example of atmospheric muons. 
The half life of a muon is 1.5 $\mu$s in its rest frame.  
A fraction 1/8 of the muons---created at 60 km above Earth 
and coming vertically down---survive at sea 
level.\cite{nalini}  Proper time between the event of 
creation of the muons and the event of muons reaching 
sea level can be measured from the muon's rest frame. 
This proper time is 4.5 $\mu$s (1.5 $\mu$s $\times$ 3).  
From the muon's rest  
frame, the proper distance between these two events 
cannot be measured.  But proper distance between 
the events can be measured from Earth and is 
60 km.  Now we can combine these two reference 
frames to construct a hypothetical reference frame by 
plotting the proper time versus the proper distance 
of the events, similar to constructing a vector from
two orthogonal vectors.  Let us call this hypothetical 
frame the proper frame. Now if we define a hypothetical 
``proper speed'' by dividing the proper distance by 
proper time,  this proper speed doesn't always have 
to be less than the conventional speed of light.  
As a matter 
of fact, the proper speed of the atmospheric muons 
is 1.3 $\times$ 10$^{10}$ m/s, over forty times faster than 
the conventional speed of light!  There is no length 
contraction 
or time dilation in the proper reference frame.  It is 
our inability to construct such a proper reference 
frame that produces all of the counterintuitive 
measurement results.  At low speed the measurement 
of the coordinate time and distance have values 
close to the proper time and proper distance, hence 
at low speed, a reference frame resembles a proper 
reference frame.

\section{CONCLUSION}
The theory of special relativity can be simplified 
conceptually by a postulate of four dimensional spacetime.  
The four dimensional spacetime postulate 
provides a geometrical view of space-time, a better logical foundation, and 
a consistent picture with the theory of general relativity.  With this 
geometrical picture, one can make a transition from general relativity 
to special relativity by simply setting the spacetime curvature equal to
zero.  All of the counterintuitive notions, including the frame 
independence of the speed of light, appear as consequences of the postulate.

In this geometrical picture of space-time, the metric equation, 
not the Lorentz transformation equations,   
is the most important equation and the most important concept.

\end{document}